\documentclass[12pt,preprint]{emulateapj}
\def\apj{{ApJ}}
\def\mnras{{ MNRAS}}

\def\gr{$\gamma$-ray}
\def\grs{$\gamma$-rays}
\def\be{\begin{equation}}
\def\ee{\end{equation}}
\def\bea{\begin{eqnarray}}
\def\eea{\end{eqnarray}}

\usepackage{graphicx}

\begin{document}

\title{High energy emission of GRB 130821A: constraining the density profile of the circum-burst medium as well as the initial Lorentz factor of the outflow}

\author{Yun-Feng Liang\altaffilmark{1}, Bei Zhou\altaffilmark{1,3}, Hao-Ning He\altaffilmark{1}, Pak-Hin~Thomas~Tam\altaffilmark{2}, Yi-Zhong Fan\altaffilmark{1}, and Da-Ming Wei\altaffilmark{1}}
\altaffiltext{1}{Key laboratory of Dark Matter and Space Astronomy, Purple Mountain Observatory, Chinese Academy of Sciences, Nanjing 210008, China.}
\altaffiltext{2}{Institute of Astronomy and Department of Physics, National Tsing Hua University, Hsinchu 30013, Taiwan.}
\email{phtam@phys.nthu.edu.tw (PHT)}
\altaffiltext{3}{Graduate University of Chinese Academy of Sciences, Yuquan Road 19, Beijing, 100049, China.}

\begin{abstract}
 GRB 130821A was detected by Fermi-GBM/LAT, Konus-Wind, SPI-ACS/INTEGRAL, RHESSI and Mars Odyssey-HEND. Although the data of GRB 130821A are very limited, we show in this work that the high energy \gr~emission (i.e., above 100 MeV) alone imposes tight constraint on the density profile of the circum-burst medium as well as the initial Lorentz factor of the outflow. The temporal behavior of the high energy \gr~emission is consistent with the forward shock synchrotron radiation model and the  circum-burst medium likely has a constant-density profile. The Lorentz factor is about a few hundred, similar to other bright GRBs.
\end{abstract}


\setlength{\parindent}{.25in}

\section{Introduction}
Gamma-ray Bursts (GRBs) are short/brief intense flashes of soft $\gamma$-rays, which have fascinated astronomers
and astrophysicists since their unexpected discovery
in the 1960s \citep{1973ApJ...182L..85K}. Their physical
origin has been debated for a long time mainly due to
the lack of an exact position and a reliable estimate of
the distance to us. In 1997, quite a few GRBs were accurately localized by the BeppoSAX
satellite, leading to the discovery of their X-ray, optical
and radio counterparts, and their redshifts \citep{1997Natur.387..783C, 1997Natur.386..686V, 1997Natur.389..261F}. The
cosmological origin of GRBs was thus directly confirmed. In the standard fireball model, the fit of the optical/radio/X-ray afterglow emission plays a key role in constraining the physical parameters including the forward shock parameters, the density profile of the circum-burst medium, the initial Lorentz factor, and sometimes even the structure, of the outflow \citep[e.g.,][]{panaitescu2001,panaitescu2002, huang2004, molinari2007, jin2007}.

Since the launch of the Fermi satellite in 2008, more than 30
GRBs have been detected above 100 MeV by
the Large Area Telescope (LAT) on board the satellite \citep{Abdo2009,Zhang2011,tam2012,lat_grb_cat}, confirming the long-lived MeV--GeV emission (i.e., longer than the prompt soft \grs) of GRBs first revealed by EGRET~\citep{Hurley94}.

One of the most surprising findings in the Fermi-era may be that the long-lived MeV-GeV emission is likely dominated by the synchrotron radiation (see Fig.3 of \citet{Zou2009} for a pioneering prediction and see e.g., \citet{Kumar2009, Gao2009, Ghisellini2010, Ackermann2013} for modeling the Fermi-LAT data in such a way) rather than the widely believed inverse Compton radiation of the forward shock electrons \citep[see, e.g.,][for a review on various inverse Compton processes]{FP08}. In contrast, the first clear evidence of an inverse Compton component of a GRB was found only recently~\citep{Fan_130427a, Tam_130427a}. Within the synchrotron radiation model, the long-lived MeV-GeV emission alone can impose tight constraint on some physical parameters. This is very important since for some bursts only the Fermi GBM/LAT-data are available. As demonstrated in this work, GRB 130821A is such an example.

This work is structured as the follows. In section 2 we briefly introduce GRB 130821A. In section 3 we present our data analysis results of the Fermi GBM and LAT data. In section 4 we discuss the implications of the data.

\section{GRB 130821A}
At 16:10:28.011 UT ($T_\mathrm{0}$) on August 21 2013, the Fermi Gamma-Ray Monitor triggered on GRB~130821A~\citep[trigger 398794231;][]{jenke15113}, which resulted in an Autonomous Repoint Request (ARR) and the LAT slewed to the GBM position. The best LAT on-ground location was reported to be R.A.~$=314\fdg1$, Dec.~$=-12\fdg0$ (J2000) with an error radius of 0$\fdg$1 \citep[68\% containment, statistical error only;][]{kocevski15115}.
The GBM light curve showed a multiple-peaked structure with a duration ($T_{90}$) of about 84s (50-300 keV).

The angle of the GRB position was about 37$^\circ$ from the LAT boresight at $T_\mathrm{0}$ and the ARR brought the source within the LAT field of view for the next 2400 seconds, so the LAT observed the GRB position with good sensitivity over the entire prompt phase. Multi-peaked emission lasting roughly 40 seconds can be seen using the non-standard LAT Low Energy (LLE) photons, which are dominated by 30--100~MeV \grs, with a significance of ~13 $\sigma$~\citep{kocevski15115}.

Konus-Wind also triggered on GRB~130821A~\citep{Golenetskii15125}. The Konus-Wind team reported that the 20~keV to 18~MeV time-averaged spectrum from 0~s to 78.08~s after the Konus-Wind trigger time is best fitted by the Band function with $\alpha=-1.33\pm0.11$, $\beta=-2.25\pm0.19$, and $E_\mathrm{p}=260\pm47$~keV; emission was seen up to $\sim$9~MeV. The burst fluence is $(9.9\pm0.9)\times10^{-5}$~erg~cm$^{-2}$~\citep{Golenetskii15125}. RHESSI, INTEGRAL SPI-ACS, and Mars Odyssey-HEND also triggered on this GRB~\citep{Hurley15127}.

Upper limits in the optical and X-ray bands were drawn because no source was found~\citep{xu15124,page15123} within the LAT error circle reported in~\citet{kocevski15115}. However, we will show that the GRB location is likely outside of this reported error circle and therefore such limits do not apply to GRB~130821A.

\section{Data analysis and results}

\subsection{Joint GBM and LAT spectral analysis during the prompt phase}
We extracted both LAT and GBM data from the Fermi Science Support Center\footnote{http://fermi.gsfc.nasa.gov/ssc/data/access/}. Joint spectral fits of both LAT and GBM data were performed over the whole prompt phase and for the four time intervals listed in Table~\ref{fitresult}, using the software package RMFIT (version 4.32). Due to the small number of photons detected by the LAT, we used C-statistic rather than chi-squared statistic to fit the data. Time Tagged Event (TTE) data from the NaI detectors n6,n7 and BGO detector b1 were used to make spectral fits for GBM, and the LAT data contains TRANSIENT class photons from 100~MeV to 300~GeV within 10$^\circ$ around the localization of (314$\fdg$27, $-11\fdg70$), while LLE data were also included. All spectra are well fitted by the Band function~\citep{1993ApJ...413..281B}, and we report the best fit model parameters in Table~\ref{fitresult}. We found that adding a blackbody component did not improve the fits.

\begin{figure*}
\includegraphics[angle=0,scale=0.350,width=0.45\textwidth,height=0.33\textheight]{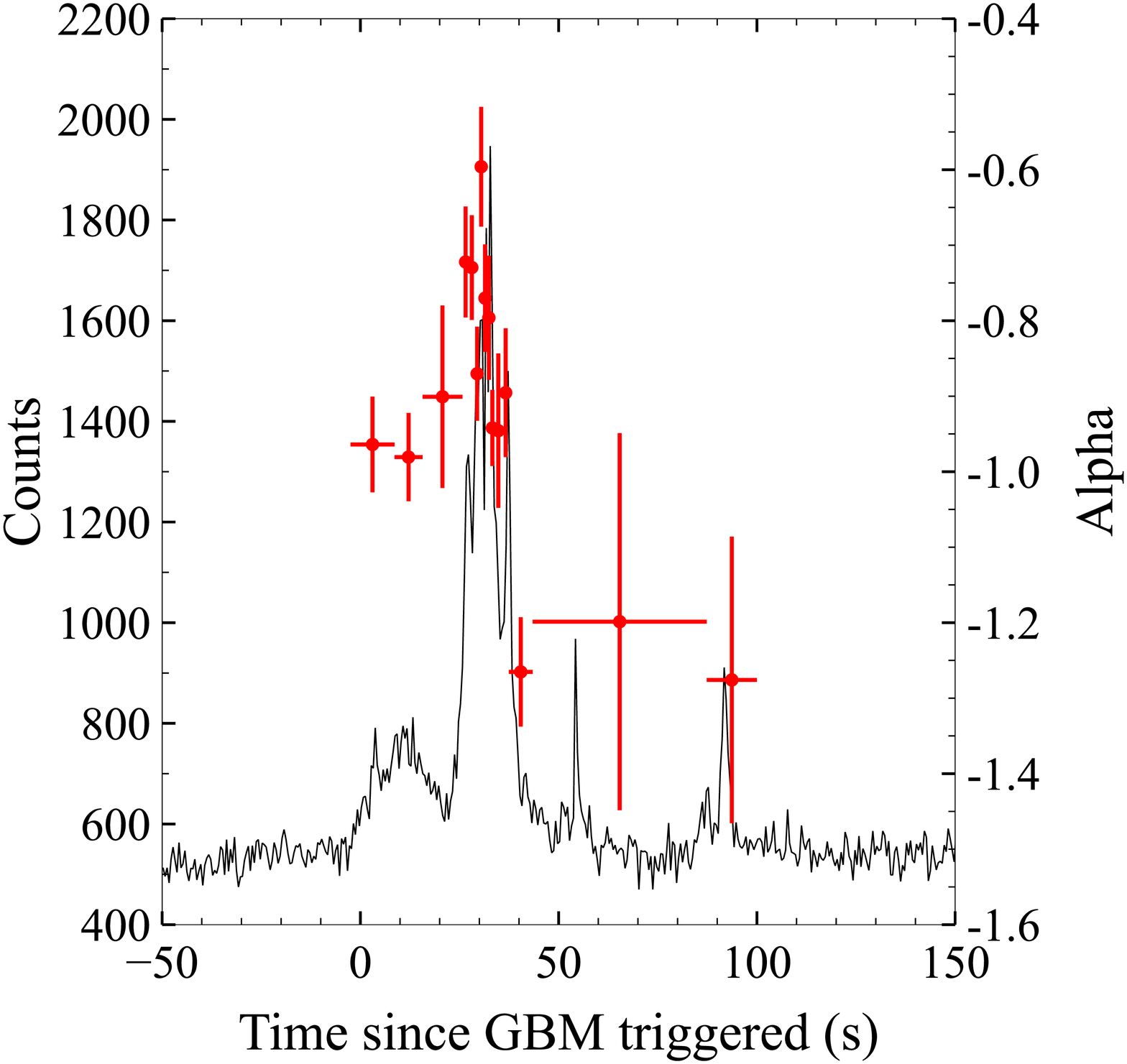}%
\includegraphics[angle=0,scale=0.350,width=0.45\textwidth,height=0.33\textheight]{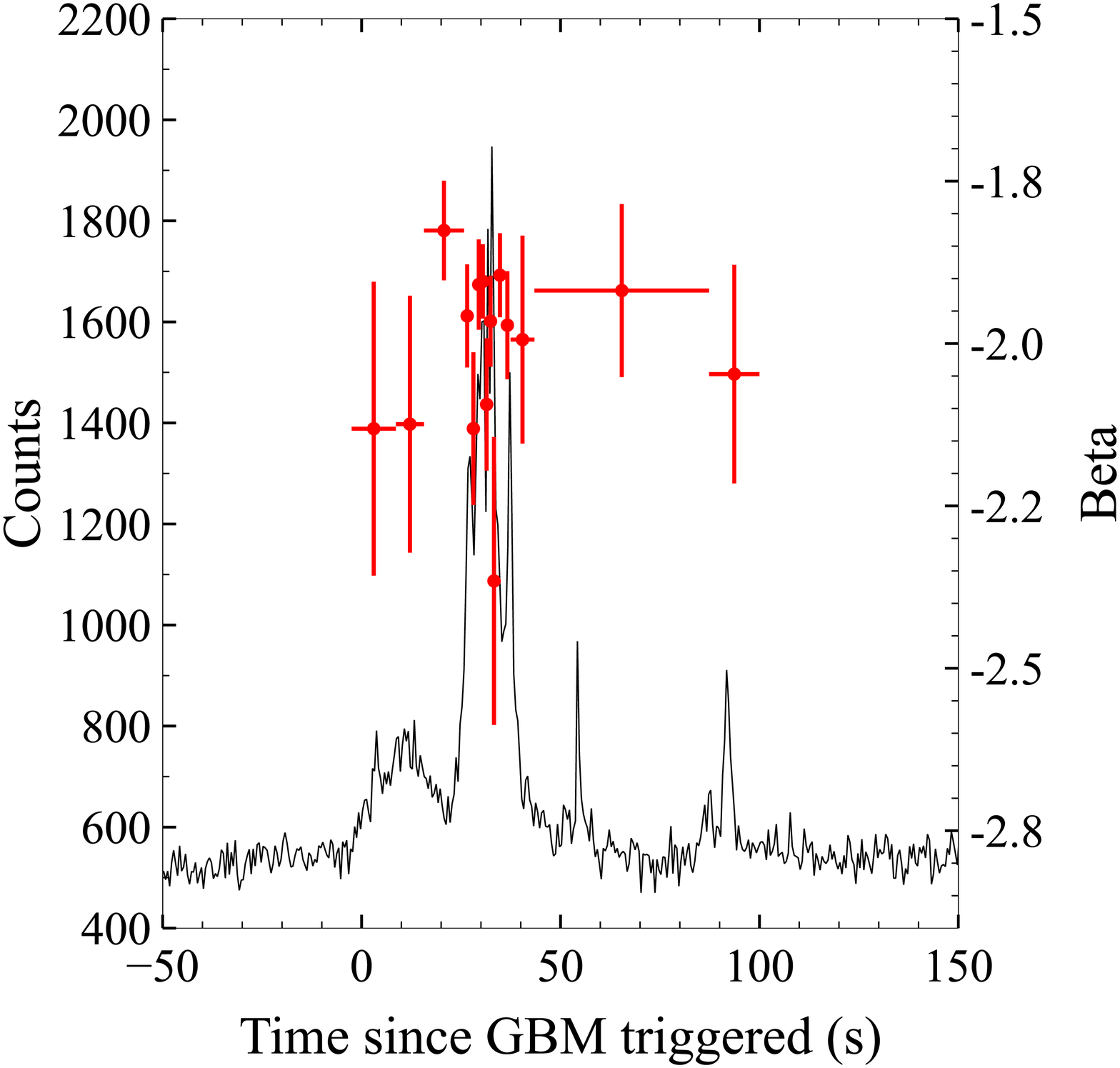}%
\\
\includegraphics[angle=0,scale=0.350,width=0.45\textwidth,height=0.33\textheight]{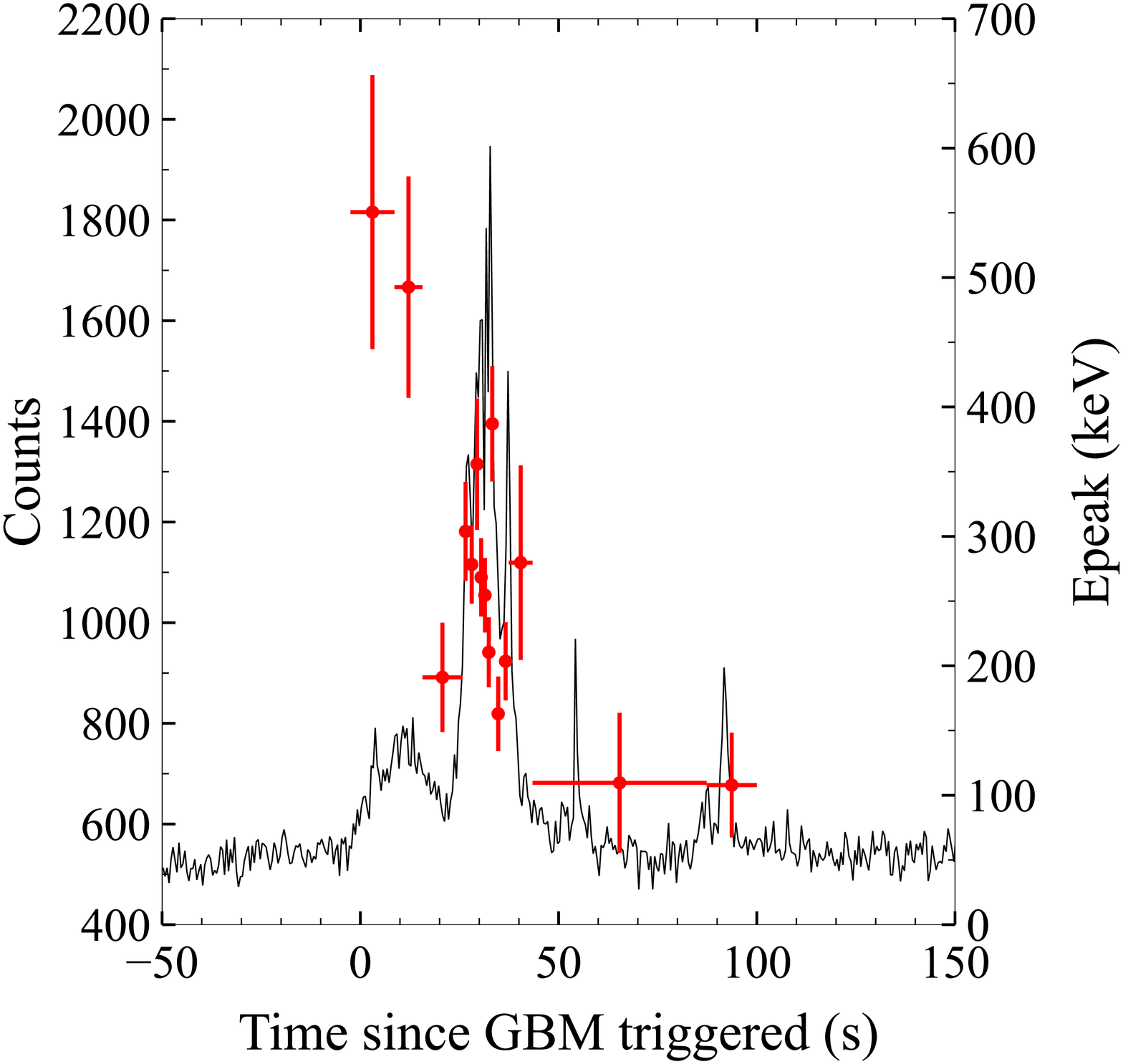}
\hfill
\caption{Evolution of the spectral parameters of the Band function with time (red point, left: $\alpha$, middle: $\beta$, right: $E_{\rm peak}$), overlaid on the light curve (black line) of GRB130821A. The light curve was extracted using the data of NaI detector n7.}
\label{batchfit}
\end{figure*}

To better understand how spectral parameters evolve with time, the best-fit values of spectral parameters derived from GBM data in different intervals are shown in Fig.~\ref{batchfit}. From the figure, there is a trend that $\alpha$ increases as the flux of the burst, while change in $\beta$ is marginal. The $E_{\rm peak}$ value gradually becomes smaller with time, which is consistent with a hard-to-soft pattern~\citep{lu2012}.

\subsection{LAT data analysis during the prompt and afterglow phases}
\label{lat_analysis}
\begin{figure*}
\centering
\includegraphics[angle=0,scale=0.350,width=0.50\textwidth,height=0.4\textheight]{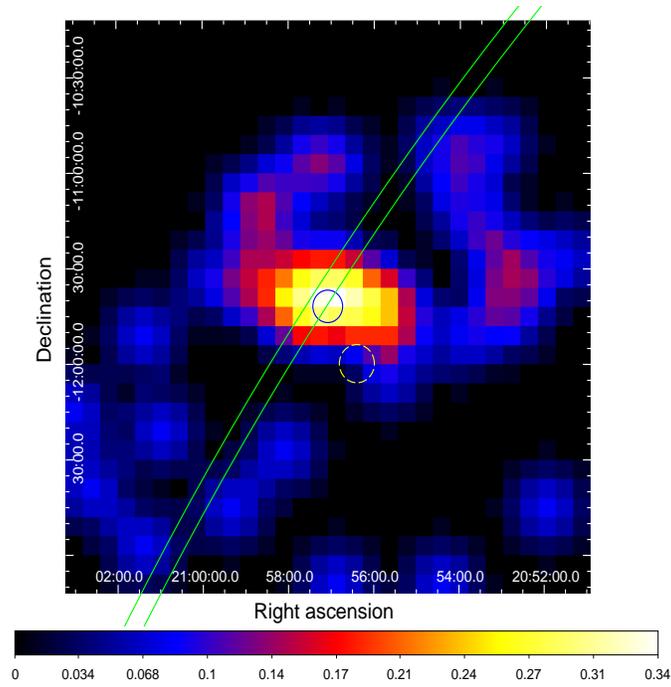}
\hfill
\caption{The 3$^\circ\times$3$^\circ$ count map of 100~MeV--20~GeV P7TRANSIENT photons that arrived between $T_{0}+0$~s and $T_{0}+1400$~s, smoothed with a Gaussian of width 0$\fdg$3. The green annulus is the GRB position derived by the IPN triangulation~\citep{Hurley15127}. The blue circle and the dashed, yellow circle shows the error circle as obtained in Sect.~\ref{lat_analysis} and~\citet{kocevski15115}, respectively. The GBM localization is outside of this field.}
\label{lat_cmap}
\end{figure*}

The Fermi Science Tools v9r31p1 package was used to analyze the data. To filter out the Earth's limb emission, we excluded events with zenith angles greater than 100 degrees in our analysis. We also used \emph{gtfindsrc} to find the best-fit position of this burst. When doing this, ``P7SOURCE'' data in time interval from $T_0+0$~s to $T_0+1400$~s of energies between 100~MeV and 300~GeV from a region of interest (ROI) of a 10$^\circ$-radius circular region centered on RA = 314$\fdg$1, Dec = $-$12$\fdg$0 (J2000)\citep{kocevski15115} were selected. The derived best location is RA = 314$\fdg$27, Dec = $-$11$\fdg$70 (J2000) with an error radius of 0$\fdg$085. Following~\citet{lat_grb_cat}, we produced a \emph{test-statistic} (TS) map (here we choose 0$\fdg$05 grid) and the maximum in the TS map is located at RA = 314$\fdg$24, Dec = $-$11$\fdg$68 (J2000) with an error radius of 0$\fdg$1, consistent with the result obtained by \emph{gtfindsrc}. We found that our derived GRB position is well consistent with the InterPlanetary Network (IPN) annulus \citep{Hurley15127}, suggesting this position is more accurate than the one reported in~\citet{kocevski15115}. A count map, using the 100~MeV to 20~GeV ``P7TRANSIENT'' data in the time interval from $T_{0}+0$~s to $T_{0}+1400$~s, is shown in Figure~\ref{lat_cmap}. The map illustrates that the LAT data that we see are indeed positionally consistent with the IPN annulus and that the LAT photons are associated with GRB~130821A. Therefore, we will use (314$\fdg$27, $-$11$\fdg$70) as the burst location in the following analysis.

\begin{figure*}
\includegraphics[angle=0,scale=0.350,width=0.50\textwidth,height=0.4\textheight]{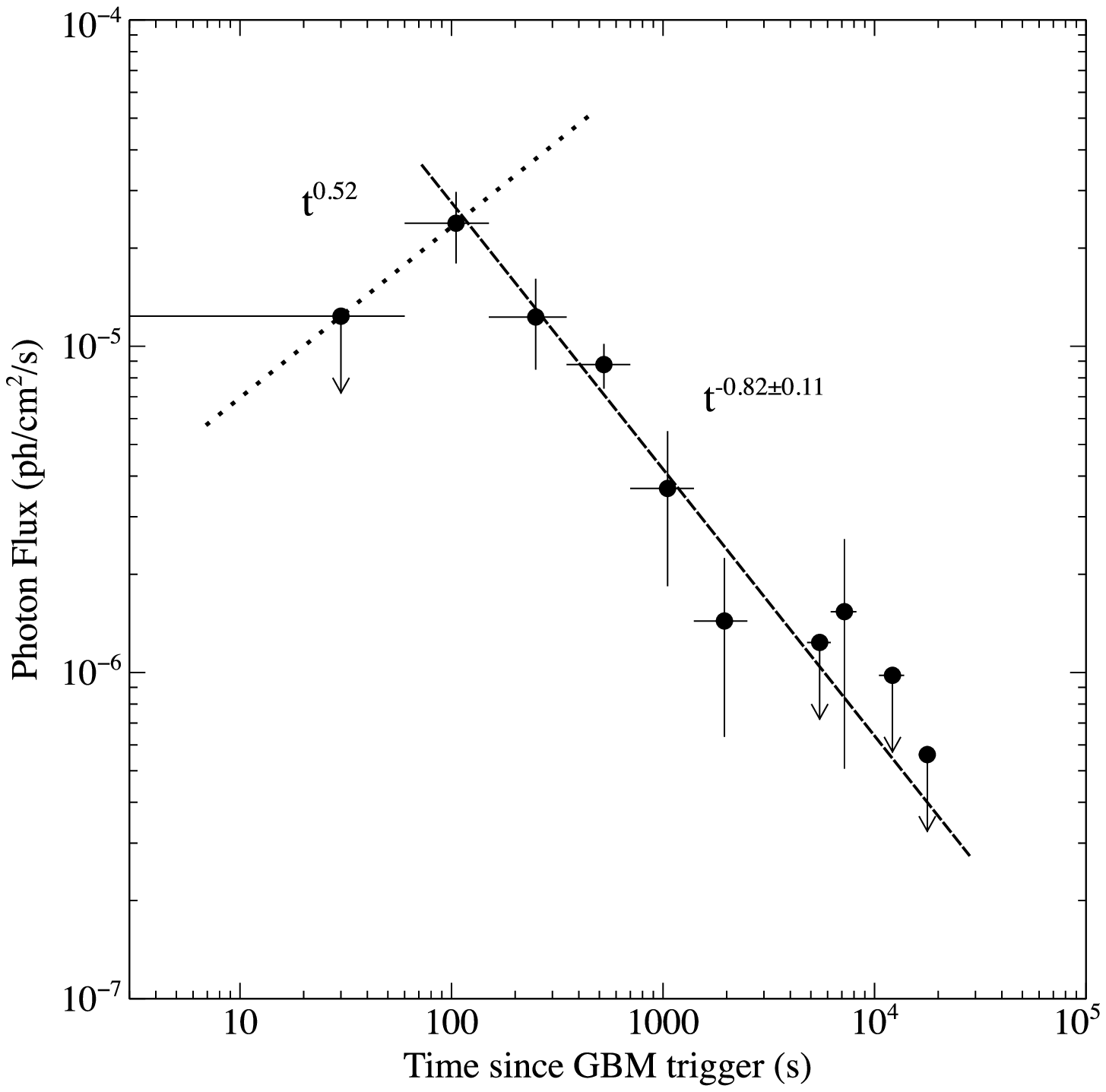}%
\includegraphics[angle=0,scale=0.350,width=0.50\textwidth,height=0.4\textheight]{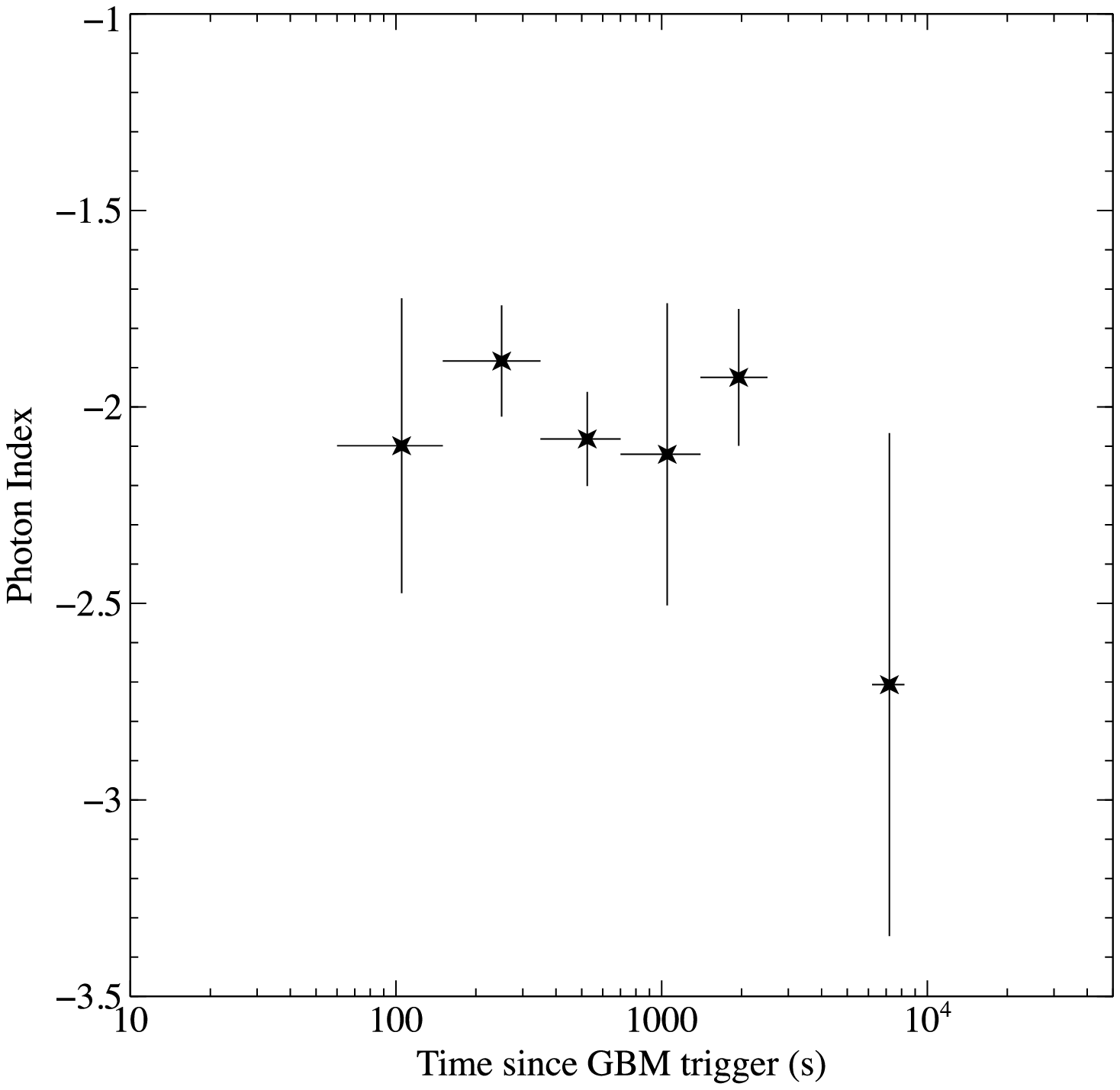}
\hfill
\caption{LAT flux(left panel) and photon index(right panel) of GRB 130821A derived from unbinned likelihood analysis. Extended emission of this burst lasted up to $\sim$8000s after the prompt emission ended. Except for the first 60s after the burst onset, the flux decays following a power law with index -0.82$\pm$0.11.}
\label{likelc}
\end{figure*}

We performed unbinned maximum-likelihood analyses to construct a light curve of GRB~130821A in the energy range in 100~MeV to 100~GeV during the time range of $T_{0}$ to $T_{0}+37000$~s. Events that are classified as P7SOURCE in a 10$^\circ$-radius circular region centered on RA = 314$\fdg$27, Dec = $-$11$\fdg$70 (J2000) were used. To subtract the background contribution, we add the ``Galactic'' (gal\_2yearp7v6\_v0) and the ``Extragalactic'' (iso\_p7v6source) diffuse components and all point sources within the ROI into our model file (generated using \emph{make2FGLxml}\footnote{http://fermi.gsfc.nasa.gov/ssc/data/analysis/user/make2FGLxml.py}). Since photon number in these short timescales is small, only parameters of 2 point sources within 5$^\circ$ from the ROI center were allowed to vary, while parameters of other point sources were fixed. Our result is shown in Fig.~\ref{likelc} and Table~\ref{likeresult}. Some features can be found in this light curve: (I) during the first 60 seconds of the prompt emission, the TS value is consistent with zero, indicating that no emission was detected, (II) from 60~s to $\sim$2500~s after the GRB onset, the flux decays as a simple power law, as has been observed for most other Fermi LAT GRBs, (III) in the time interval 6200s--8200s, the spectrum is softer than earlier ones, but they are consistent within the uncertainties. We fit the light curve with a simple power law for the time bins with TS$>$9, and found the slope of the best-fit line to be $-$0.82$\pm$0.11, which is shown as the solid line in Fig.\ref{likelc}. If the last flux point (i.e., 6200s--8200s) is not included, the decay index is $-0.84\pm0.13$, whicis is consistent with the above value.

\begin{figure*}
\centering
\includegraphics[angle=0,scale=0.350,width=0.6\textwidth,height=0.45\textheight]{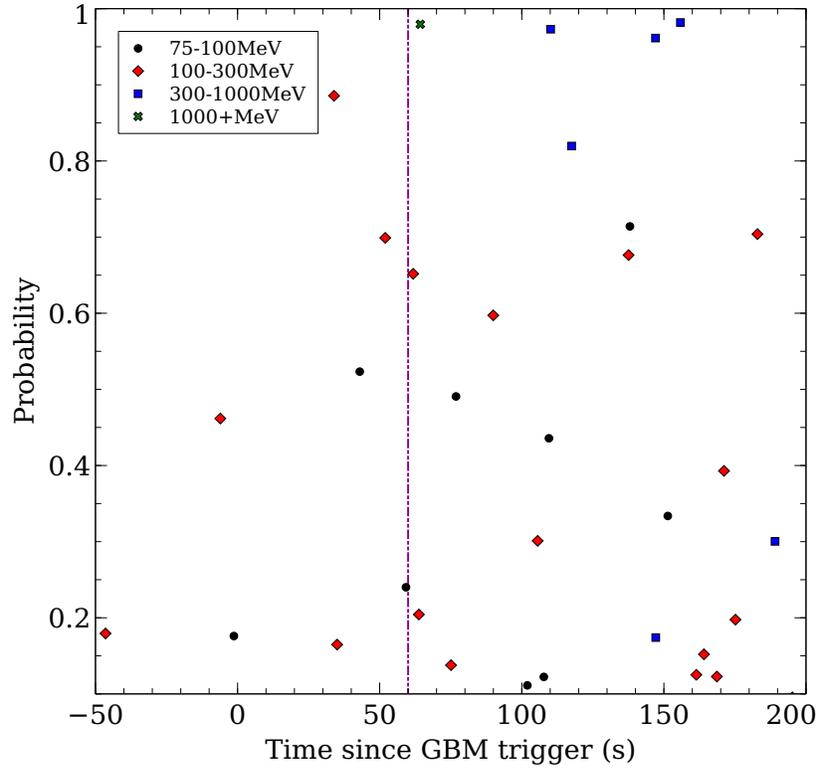}
\hfill
\caption{Probability of a photon being associated with GRB130821A and its arrival time. Photons are shown with four colors and shapes, each indicating one energy range. The vertical dashed line indicates the time $T_0+60$s. We can see that only two $>$100MeV photons with a probability$>$0.5 appeared in the first 60s. During the prompt phase, the most energetic photon arrived at $T_0+$64.3s with energy 2.7~GeV.}
\label{prob}
\end{figure*}

An interesting point we note from Fig.\ref{likelc} is that there is no significant $>$100 MeV emission up to $\sim$$T_0+$60s. To verify this, we calculate probabilities of each photon being associated to GRB 130821A by using the Fermi ScienceTool \emph{gtsrcprob}, assuming a photon index of $\Gamma=-$2.1, and plot the probabilities v.s. photons' arrival times in Fig.\ref{prob}. All those TRANSIENT class events in a 30$^\circ$-radius circular region centered on (314$\fdg$27, $-11\fdg70$) were considered, but only events with a probability being associated to GRB~130821A greater than 0.1 are plotted. Only 2 photons with probabilities larger than 0.5 arrived in the first 60 seconds of the prompt emission, energies of both of which are less than 300~MeV.

The fact that there is no significant detection at $>$100~MeV by the LAT during the first 60s after the GRB onset, during which most of the prompt MeV emission was seen, and $E_{\rm peak}$ was the highest, suggests that (I) the $>$100~MeV emission is not an extrapolation to the Band function during the prompt phase, and (II) the LAT and the GBM emission evolve independently. In turn, GRB~130821A may be the first-ever GRB where most or all $>$100~MeV emission is unrelated to the prompt emission. The $>$100~MeV emission must have a different origin than the prompt emission.

\section{Discussion}
As shown in Fig.\ref{batchfit}, most of the prompt emission concentrated in the first 40 seconds after the burst onset. However, one can see from Figs.~\ref{likelc} and \ref{prob} that very rare $>100$~MeV photons have been detected in such an interval (see also Table~\ref{tab:like_fit}) and the most energetic $\gamma$-ray at an energy $\sim$6~GeV arrived at 219~s after the trigger. The Fermi-LAT data likely peaked at $t_{\rm p}\sim 100$s, thus lagging behind the soft $\gamma-$ray peak emission significantly. After the high energy peak emission, the count rate drops with time as $t^{-0.82\pm0.11}$. All these behaviors are consistent with the forward shock synchrotron radiation model. Below we take such a model and show that some interesting results can be achieved.

The rising behavior of the forward shock synchrotron radiation light-curve sheds valuable light on the density profile of the circum-burst medium \citep[see Table 1 in][for a summary]{xue2009}. For GRB 130821A, the high energy emission occurred significantly after the strongest soft gamma-ray emission phase had ended, hence the GRB outflow should be in the thin shell regime for which $t_{\rm p}>T_{90}$, where $t_{\rm p}$ is the time when reverse shock crosses the ejecta), as defined in \citet{xue2009}. For $t<t_{\rm p}$ and the number density of the medium $n\propto R^{-k}$ ($k=0$ for interstellar medium and $=2$ for free stellar wind), we have the typical synchrotron radiation (cooling) frequency $\nu_{\rm m} \propto t^{-k/2}$ ($\nu_{\rm c} \propto t^{3k/2-2}$) and the maximal specific flux $F_{\rm \nu,max} \propto t^{3(1-k/2)}$.
For typical forward shock parameters, the observer's frequency $\nu_{\rm obs}=100$ MeV is well above both \footnote{At $t\geq t_{\rm p}$, the  $\nu_{\rm m}$ and $\nu_{\rm c}$ are estimated with Eqs.~(1) and (2) of Yost et al. (2003). For $t<t_{\rm p}$, we have $\nu_{\rm m}(t)=(t/t_{\rm p})^{-k/2}\nu_{\rm m}(t_{\rm p})$ and $\nu_{\rm c}(t)=(t/t_{\rm p})^{3k/2-2}\nu_{\rm c}(t_{\rm p})$. Then it is straightforward to show that with typical parameters $E_{\rm k} \sim 10^{54}$ erg, $\epsilon_{\rm e}<1/3$, $\epsilon_{\rm B} \sim 0.01$ and $p\sim 2.3$, we have $\max\{\nu_{\rm m},\nu_{\rm c}\}\ll 100~{\rm MeV}$ for $t \sim 10-100$ s. \label{footnote-1}} $\nu_{\rm c}$ and $\nu_{\rm m}$  and we have $F_{\nu_{\rm obs}} \propto t^2$ ($\propto t^{(2-p)/2}$) for $k=0$ ($=2$), where $p>2$ is the power-law index of shocked electrons, as suggested by the simultaneous Fermi-LAT spectrum. Thus, in order to reproduce the quick rise (quicker than $t^{1/2}$, see Fig.\ref{likelc}) of the high energy emission for $t<100$ s, the particle density of the circum-burst medium should be a constant. 
The stellar wind medium model might be able to marginally match the data if the typical synchrotron radiation frequency of the forward shock $\nu_{\rm m}> 100$ MeV at $t\sim 100$ s (see Table~1 of Xue et al. 2009; for current discussion, one should replace $\nu_{_{\rm X}}$ therein by $\nu_{\rm obs}$). However, as already mentioned in the footnote \ref{footnote-1}, in the standard fireball model, for reasonable parameters, $\nu_{\rm m}$ is expected to be  $\ll 100$ MeV, so is $\nu_{\rm c}$. 
For $t>t_{\rm p}$, the forward shock synchrotron radiation at energies above 100 MeV drops with time as $\propto t^{-1}$ for $p \sim 2$ and $\max\{\nu_{\rm m},~\nu_{\rm c}\}<\nu_{\rm obs}$, in agreement with the detected decline, where $p$ is the power-law index of the shock-accelerated electrons. The high energy emission is thus $F_{\nu_{\rm obs}} \propto {\nu_{\rm obs}}^{-p/2} \sim \nu_{\rm obs}^{-1}$, consistent with the data, too (see the right panel of Fig.\ref{likelc}).

In the thin shell case, the afterglow peak time traces the deceleration of the forward shock and in turn can be used to constrain the initial Lorentz factor of the GRB outflow \citep[e.g.,][]{sari1999,molinari2007}
\begin{equation}
\Gamma_0  = [\frac{24E_{\rm k}(1+z)^3}{\pi n m_p c^5 t_{\rm p}^3}]^{1/8},
\end{equation} \\
where $m_p$ represents the proton mass, $c$ the speed of light, and $t_\mathrm{p}$ the peak time of the $>$100~MeV emission, respectively. We assume the ambient density $n=1~\rm cm^{-3}$.
The isotropic energy of the outflow $E_{\rm k}$ can be estimated based
on the total energy of prompt gamma-ray emission $E_{\gamma}$
assuming a certain radiation efficiency $\eta$. For GRB 130821A, the redshift is unknown and we take the typical redshift of current GRBs, i.e., $z=1$. Using the energy fluence given by the Konus-Wind, we have $E_{\gamma}=(2.5{\pm}0.2){\times}10^{53}$~erg. We take $\eta=0.2$ in the calculation according to~\citet{guetta2001}. As a result, we get $\Gamma_0\sim440$.

The initial bulk Lorentz factor can be estimated in an alternative way. In the forward shock synchrotron model, it is widely known that the maximal  radiation frequency can be estimated as (e.g., Cheng \& Wei 1996)
\begin{equation}
\epsilon_{\rm M} \sim 100~{\rm MeV}~\Gamma/(1+z).
\end{equation}
Therefore the fact that the highest energy LAT photon at an energy of $\sim6$~GeV arrived 219~s after the GRB trigger suggests an initial Lorentz factor
$\Gamma_0 \gtrsim 200[(1+z)/2]$, where the temporal decay of the Lorentz factor $\Gamma \propto (t/t_{\rm p})^{-3/8}$ has been taken into account. Interestingly, the result, i.e., $\Gamma_0 \sim 350$), is consistent with the independent constraint using the forward shock deceleration argument.

In view of these facts, we conclude that: (i) the high energy emission of GRB 130821A may indeed have a forward shock synchrotron radiation origin; (ii) the circum-burst medium likely has a constant density profile; (iii) the outflow is jetted and ultra-relativistic with an initial Lorentz factor of a few hundred. Our results demonstrate that the long-lived MeV-GeV emission alone can impose tight constraints on some physical parameters.

\section*{Acknowledgments}
We thank H. F. Yu for the helpful discussion on GBM analysis of this burst. This work is supported in part by 973 Program of China under grant 2013CB837000, National Natural Science of China under grants 11163003 and 11273063, and by China Postdoctoral science foundation under grant 2012M521137. YZF is also supported by the 100
Talents program of Chinese Academy of Sciences and the Foundation for
Distinguished Young Scholars of Jiangsu Province, China (No. BK2012047). PHT is supported by the National Science Council of the Republic of China (Taiwan) through grant NSC101-2112-M-007-022-MY3.

\begin{deluxetable}{ccccc@{}r@{}c@{}c}
\tablewidth{0pt}
\tabletypesize{\footnotesize}
\tablecaption{Summary of GBM/LAT Joint Spectral Fitting Using Band Function.\label{fitresult}}
\tablehead{
\colhead{} & \colhead{Interval} & \colhead{$E_\mathrm{peak}$} & \colhead{$\alpha$} & \colhead{$\beta$} & \colhead{Photon Flux} & \colhead{Energy Flux} & \colhead{C-stat/dof} \\
\colhead{} & \colhead{(sec)} & \colhead{(keV)} & \colhead{} & \colhead{} & \colhead{(cm$^{-2}$s$^{-1}$)} & \colhead{(10$^{-7}$erg~cm$^{-2}$s$^{-1}$)} & \colhead{}
}
\startdata
All& $-$2.0--100.1&374.4$\pm$15.0&$-$1.10$\pm$0.02&$-$2.75$\pm$0.04&4.30$\pm$0.03&7.14$\pm$0.08&830/391\\
a& $-$2.0--21.0&588.5$\pm$55.7&$-$1.06$\pm$0.03&$-$3.11$\pm$0.18&4.00$\pm$0.06&8.34$\pm$0.17&520/391\\
b&21.0--46.6&332.6$\pm$10.5&$-$0.98$\pm$0.02&$-$2.75$\pm$0.05&10.45$\pm$0.07&17.9$\pm$0.20&630/391\\
c&46.6--66.3&151.2$\pm$33.3&$-$1.22$\pm$0.13&$-$2.33$\pm$0.06&1.69$\pm$0.07&1.82$\pm$0.09&441/391\\
d&80.4--100.1&111.1$\pm$19.5&$-$1.21$\pm$0.14&$-$2.34$\pm$0.06&2.00$\pm$0.07&1.91$\pm$0.08&423/391\\
\enddata
\label{Band_fit}
\end{deluxetable}

\begin{deluxetable}{ccccc}
\tablewidth{0pt}
\tablecaption{LAT likelihood analysis result using power law.\label{likeresult}}
\tablehead{
\colhead{Time(s)} & \colhead{Flux\tablenotemark{a}} & \colhead{Photon Index\tablenotemark{b}} & \colhead{TS\ value} & \colhead{Npred\tablenotemark{c}}
}					
\startdata
0--60&1.24$\times$10$^{-5}$& - &0.0&0.0\\
60--150&(2.51$\pm$1.08)$\times$10$^{-5}$&$-$2.1$\pm$0.4&32.66&6.75\\
150--350&(1.19$\pm$0.23)$\times$10$^{-5}$&$-$1.9$\pm$0.1&40.36&9.78\\
350--700&(8.70$\pm$1.16)$\times$10$^{-6}$&$-$2.1$\pm$0.1&56.18&12.57\\
700--1400&(3.63$\pm$2.09)$\times$10$^{-6}$&$-$2.1$\pm$0.4&23.50&9.80\\
1400--2500&(1.48$\pm$0.36)$\times$10$^{-6}$&$-$1.9$\pm$0.2&22.47&5.88\\
4800--6200&1.24$\times$10$^{-6}$& - &0.0&0.0\\
6200--8200&(1.77$\pm$1.08)$\times$10$^{-6}$&$-$2.7$\pm$0.6&10.22&10.78\\
10500--13800&9.80$\times$10$^{-7}$& - &1.35&1.42\\
16200--19300&5.60$\times$10$^{-7}$& - &0.0&0.0\\
21900--25400&5.84$\times$10$^{-7}$& - &0.0&0.0\\
27500--36700&8.23$\times$10$^{-7}$& - &0.0&0.0\\
\enddata
\tablenotetext{a}{In the unit of photons~cm$^{-2}$s$^{-1}$; values without uncertainty are upper limits.}
\tablenotetext{b}{Index values are not well constrained for the time intervals with TS$<9$.}
\tablenotetext{c}{predicted photon number given by the unbinned likelihood analysis}
\label{tab:like_fit}
\end{deluxetable}

\end{document}